\documentclass[aps,prl,preprint,groupedaddress,amsmath,amssymb]{revtex4-1}

\usepackage{natbib}
\usepackage{graphicx}
\usepackage{tabu}

\begin{document}


\title{Direct observation of charge mediated lattice distortions in complex oxide solid solutions}


\author{Xiahan Sang, Everett D. Grimley, Changning Niu, Douglas L. Irving, James M. LeBeau}
\email[]{jmlebeau@ncsu.edu}
\affiliation{Department of Materials Science \& Engineering, North Carolina State University}


\date{\today}

\begin{abstract}
Material properties depend sensitively on picometer scale atomic displacements introduced by local chemical fluctuations.  Direct real-space, high spatial-resolution measurements of this compositional variation and corresponding distortion can provide new insights into materials behavior at the atomic scale. Using aberration corrected scanning transmission electron microscopy combined with advanced imaging methods, we observed atom column specific, picometer-scale displacements induced by local chemistry in a complex oxide solid solution.  Displacements predicted from density functional theory were found to correlate with the observed experimental trends. Further analysis of bonding and charge distribution were used to clarify the mechanisms responsible for the detected structural behavior. By extending the experimental electron microscopy measurements to previously inaccessible length scales, we identified correlated atomic displacements linked to bond differences within the complex oxide structure. 

\begin{description}
\item[PACS numbers]
68.37.Ma, 71.15.Mb
\end{description}

\end{abstract}



\maketitle

Complex oxides exhibit a range of properties including ferroelectricity and piezoelectricity \cite{Dawber:2005dw,Scott:2007aa}, high temperature superconductivity \cite{Reznik:2006gf}, and thermoelectricity \cite{Koumoto:2006jo}, which are intimately linked to picometer-scale atomic shifts within the crystal structure \cite{Grinberg:2002aa}. For oxide solid solutions, the interaction between atomic species can   modify functionality through these structural distortions. To date, however, these displacements are often difficult to unambiguously detect by conventional means  \cite{Billinge:2007yq}.    Rather, direct real-space measurements with picometer precision in combination with theoretical investigations can provide insights into the origin of fine scale atomic displacements. 

Atomic resolution electron microscopy has served as an indispensable tool for the real-space structural analysis of materials, e.g.~ferroelectrics \cite{Jia:2007aa,Jia:2008aa}. High-angle annular dark field scanning transmission electron microscopy (HAADF STEM) has proven pivotal for structural analysis as the image intensities scale approximately as the square of the atomic numbers (Z) present \cite{Nellist:2011aa}. This enables direct identification of light and heavy elements within a structure, even single dopant atoms \cite{Voyles:2002aa,Krivanek:2010aa,Ishikawa:2013uq}. Crystallographic analysis using STEM, however, has been stymied by sample drift and noise that traditionally limit measurement precision to about 5-10 pm \cite{Krivanek:2010aa,Kimoto:2010aa,Zuo:2014aa}. This challenge hindered the possibility to directly connect complex structural responses to local chemistry. The recent introduction of revolving STEM \cite{Sang:2014qy} and non-ridged image registration \cite{Yankovich:2014fk} have now overcome these challenges.  The picometer precise measurements enabled by these techniques are now opening a new avenue to study crystallography with the combination of interpretability, chemical sensitivity, and directly in real-space.

\begin{figure*}
     \includegraphics[width = 170 mm]{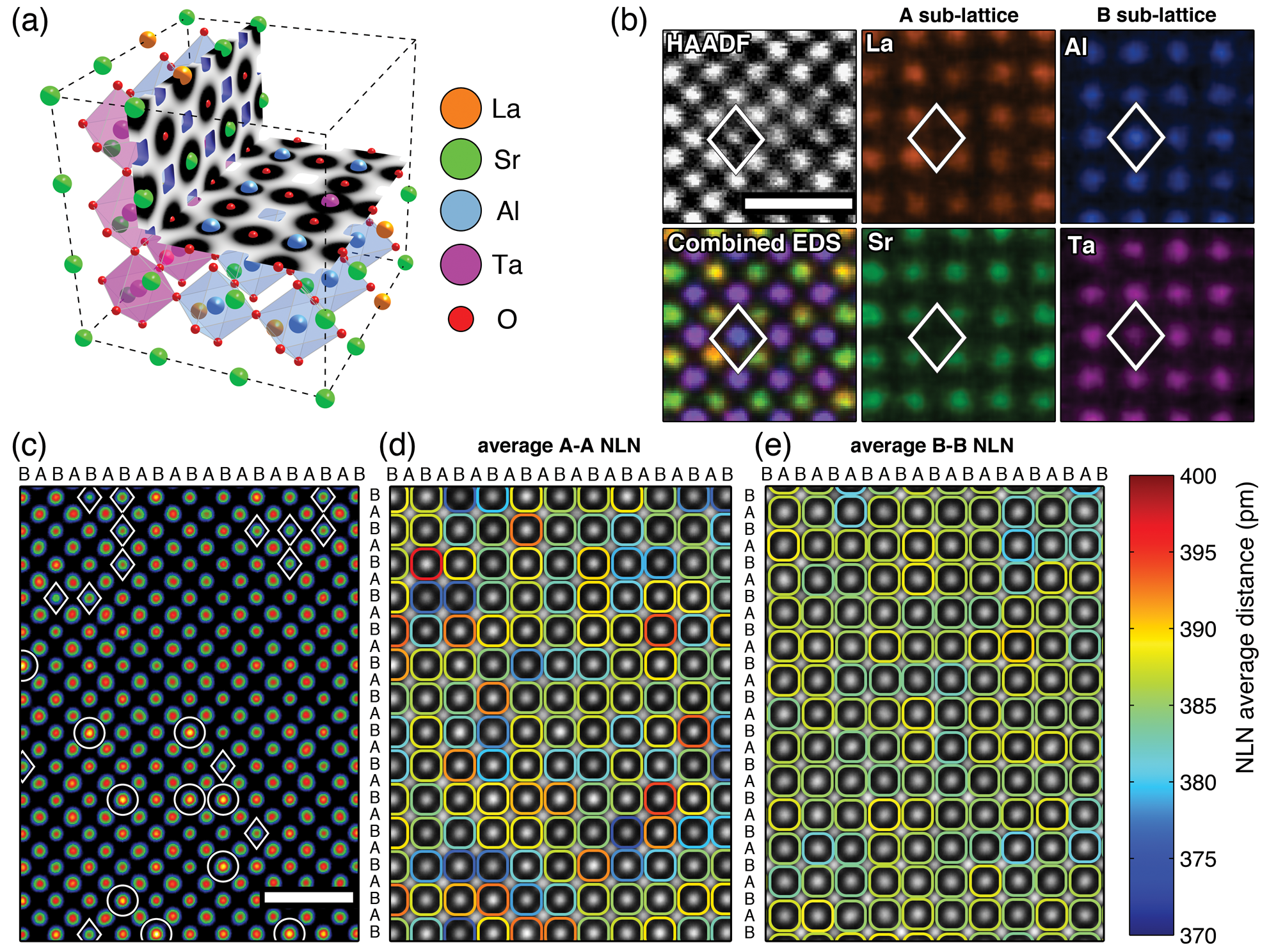}
  \caption{(a) A $3\times3\times3$ supercell of the LSAT structure superimposed with charge density for the La/Sr and Al/ Ta containing planes, vertical and horizontal slices respectively. (b) Atomic resolution energy dispersive X-ray spectroscopy (EDS) where the diamond indicates an Al rich column. (c) Sub-section of a RevSTEM image along $\left<100\right>$ with the labels `A' and `B' denoting the corresponding sub-lattices. Contrast and brightness have been adjusted to highlight the atom column intensity differences.  Strong and weak B sub-lattice atom columns are indicated by circles and diamonds respectively.  Supplemental Material \cite{Sup_Mater} Figure 1 provides the original full-size image. Average A-A (d) and B-B (e) nearest like-neighbor (NLN) distance around each sub-lattice atom column.  The indicated scale bars represent 1 nm.}
  \label{fig:figure1}
\end{figure*}

In this Article, we select (La$_{0.18}$Sr$_{0.82}$)(Al$_{0.59}$Ta$_{0.41}$)O$_3$ (LSAT) as a model oxide solid solution to study atomic displacements in a complex chemical environment.  The LSAT crystal poses a particularly challenging test case as atomic forces resulting from the intricate local chemical environment lead to structural distortion.  LSAT adopts the cubic perovskite ABO$_3$ structure (space group $Pm\bar{3}m$) with the corners occupied by (La, Sr),  (Al, Ta) located at the cube center, and the oxygen anions positioned on the faces (Fig.~\ref{fig:figure1}a).  When the LSAT structure is projected along $\left<100\right>$, chemically distinct atom columns are observed that contain either La/Sr, Al/Ta/O, or exclusively O. For convenience in referencing the $\left<100\right>$ STEM images, we define two distinct cation containing sub-lattices: the A sub-lattice containing La and Sr, and the  B sub-lattice consisting of Al and Ta with overlapping oxygen anions.

As a critical first step to identifying the underlying structure of the solid solution, measurement of the elemental distribution within LSAT is required, and is accomplished through atomic resolution energy dispersive X-ray spectroscopy (EDS) (Fig.~\ref{fig:figure1}b) \cite{DAlfonso:2010aa,Lu:2014fr,Kothleitner:2014mz}. The stronger Al X-ray signal at the darkest atom columns in the HAADF image (Fig.~\ref{fig:figure1}c) confirm that these positions are Al-rich compared to other B sub-lattice columns.  This strong contrast results from the large atomic number difference between Al (Z = 13) and Ta (Z = 73), compared to Sr (Z = 38) and La (Z = 57). To aid visual inspection, particularly strong and weak B sub-lattice atom columns are identified in Figure \ref{fig:figure1}c for those within $\pm 1.5\sigma$ of the average B column intensity.

As shown in the RevSTEM dataset (Fig.~\ref{fig:figure1}c through e), the compositional fluctuation and non-uniform lattice distortion are directly observed. This is quantified through the average nearest-like-neighbor (NLN) distances for the two distinct sub-lattices (Figs.~\ref{fig:figure1}d and e). In Figures~\ref{fig:figure1} d and e, each column is outlined by the average displacement with the extreme contraction and expansion colored as blue or red, respectively, while minor distortion is colored aqua/green/yellow.  Inspection of these maps reveals a wide distribution for the A-A NLN distances (Fig.~\ref{fig:figure1}d), while a much narrower distribution is found for the B-B NLN distances (Fig.~\ref{fig:figure1}e).  Analysis of the average A-A distances uncovers that significant deviations are correlated with the B site intensity (Fig.~\ref{fig:figure1}d).  Specifically, there is a tendency for significant contraction around the dark Al-rich atom columns  while the bright Ta-rich columns correspond to expansion. 

\begin{figure}[h!]
      \includegraphics[width = 3.5in]{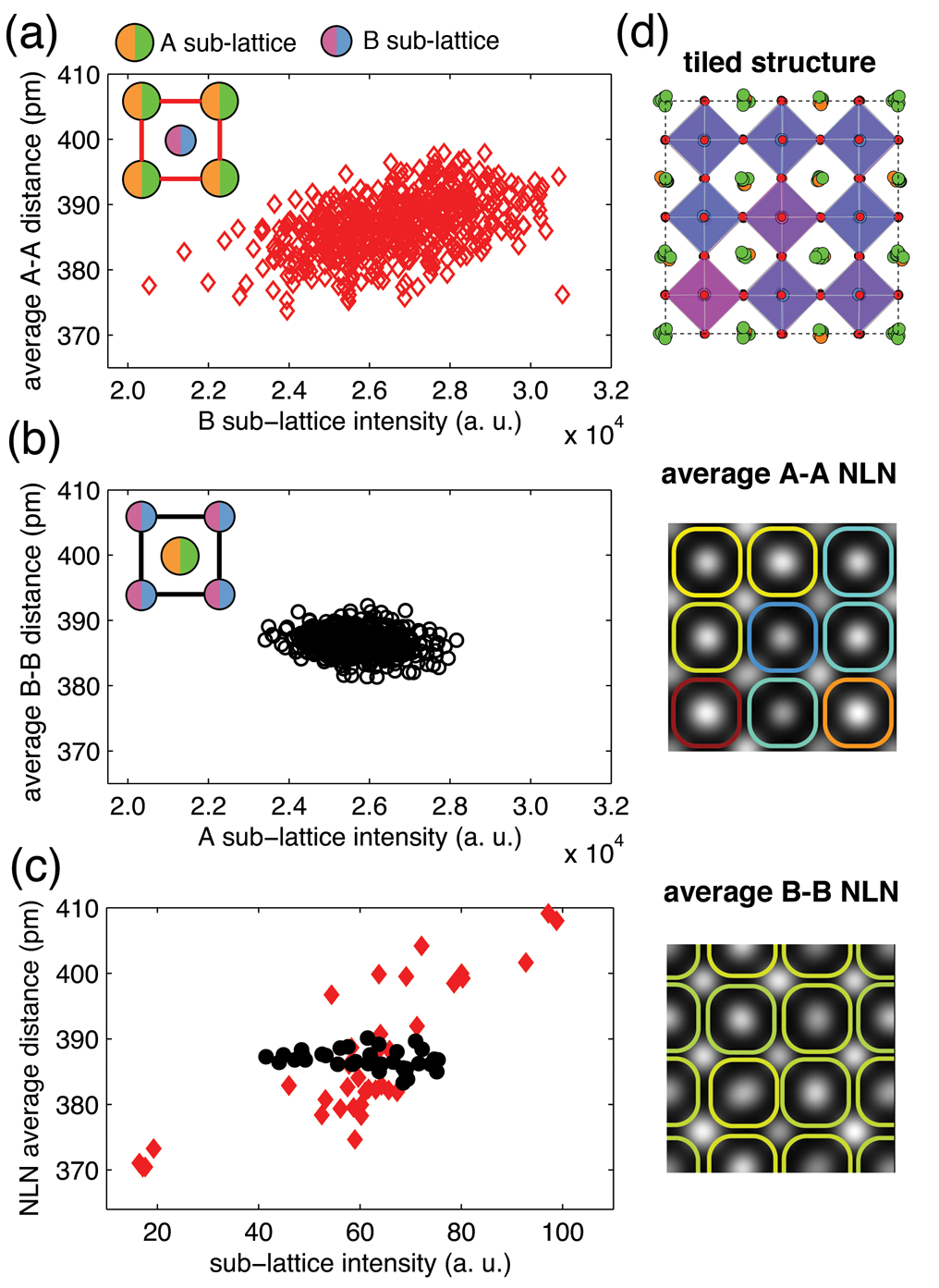}
   \caption{(a) The average A-A and (b) B-B NLN distance versus the intensity of the center, opposing sub-lattice atom column. (c) The compiled average NLN distances for both A-A and B-B versus the intensity from simulated HAADF images based on the relaxed DFT $3\times3\times3$ supercell. (d) Average NLN distances mapped across the tiled $3\times3\times9$ superstructure as in Figure \ref{fig:figure1}.}
   \label{fig:figure2}
 \end{figure}

To emphasize these trends, the average NLN distances around each B and A atom column are plotted against the opposing atom column sub-lattice intensity (Fig.~\ref{fig:figure2} a and b respectively). The average A-A distance shows moderate  linear correlation with the B atom column intensity  ($R \sim 0.5$), while the B sub-lattice distances show little to no correlation to A intensity ($R \sim 0.2$).  These trends point to the strong influence of the atomic displacements mediated by fluctuations in local chemistry. Furthermore, the scatter in these plots is a direct consequence of the statistical nature of the solid solution alloy and each measurement point represents the average local distortion.  These results are also critically relevant across a range of oxides, where previous theoretical investigations have indicated that the ferroelectricity in Pb(Ti,Zr)O$_3$ results when A sub-lattice (Pb) cations contract towards Ti and away from Zr  on the B sub-lattice \cite {Grinberg:2002aa,Grinberg:2007ab}. Similar behavior has been predicted for Bi(Mg,Ti)O$_3$ where the B sub-lattice cations have different preferred charge states and results in a ferroelectric phase transition \cite{Suewattana:2012aa}. 

In contrast to Bi(Mg,Ti)O$_3$ or Pb(Ti,Zr)O$_3$, the LSAT solid solution creates a complex local charge environment on \emph{both} cation sub-lattices.  The A sub-lattice contains La$^{3+}$ and Sr$^{2+}$, the B sub-lattice contains Al$^{3+}$ and Ta$^{5+}$, and O$^{2-}$ neutralizes the total charge. To understand the nature of bonding and the origin of forces in this complex system that drive the aforementioned distortions, we turn to density functional theory (DFT). From DFT the charge of cations in the unrelaxed structure, as measured by Bader's method \cite{Bader:1990aa, Henkelman:2006aa}, reveals that the atoms do not reside in their full ionization states. Rather,  La, Sr, Al, and Ta cations are determined to have positive charges of 2.04 $\pm$ 0.02, 1.56 $\pm$ 0.01, 2.43 $\pm$ 0.01, and 2.54 $\pm$ 0.01 $|e|$, respectively. The small standard deviations indicate that all cations are insensitive to their local chemical environment. Ta atoms have the most significant deviation from their formal charge. This is due to Ta-O bonds having the most significant covalent character as seen in the higher concentration of electrons between Ta-O atoms in Figure~\ref{fig:figure1}a. 

The oxygen anions exhibit an average charge of $-1.37 \pm 0.11$ $|e|$ but have a much larger standard deviation, which indicates a higher sensitivity to the local chemical environment. The identity of the neighboring B sub-lattice cations are found to determine the charge of oxygen.  Separating the oxygen charge into groups based on the possible combinations, the three distinct local oxygen charges are: -1.51 $\pm$ 0.01 $|e|$ for Al-O-Al, -1.33 $\pm$ 0.02 $|e|$ for Al-O-Ta, and -1.23 $\pm$ 0.02 $|e|$ for Ta-O-Ta. The small standard deviations indicate that the oxygen anions are uniquely identified within these three groups. 

The non-uniform charge distribution in the ideal lattice leads to sizable forces on the atoms of up to $\sim$4 eV/\AA, which are ultimately minimized through static displacements in the relaxed structure.  In an effort to further confirm the origin of the forces as arising from the local charge distribution, we compare the Hellmann-Feynman forces from DFT to those from Coulomb's law using the Bader charges and an Ewald summation method. Forces on atoms in the A sub-lattice derived from the Ewald sum exhibit a minimum absolute percent error (MAPE) of 7\%  relative to the DFT forces. In contrast, forces on oxygen and B sub-lattice cations have higher MAPE of 102\% and 53\%, respectively, due to mixed ionic/covalent bonding. 

The similarity of the Hellmann-Feynman forces and those from the ideal point charge Coulombic picture for the A sub-lattice cations indicate a predominant ionic interaction between these atoms and the others within the alloy.  
La and Sr are thus most sensitive to the local charge distribution, lack bond directionality, and are free to arbitrarily displace to minimize internal energy and reduce the large initial forces. Ultimately, attractive interactions arise between the A sub-lattice cations and the oxygen anions, which have variable charge based on local chemistry.  The DFT charge analysis indicates that the most favorable attraction is between A sub-lattice cations and oxygen anions interacting with Al (i.e.~Al-O-Al or Al-O-Ta).  Further, relaxation of the atomic coordinates shows a tendency for contraction of A sub-lattice cations towards Al rich B columns, consistent with the proposed mechanism and experimental measurements.

To provide analysis complementary to experiment,  STEM images were simulated using the atomic coordinates from the relaxed DFT supercell.  Two factors are included for comparison with experiment. First, the coordinates are scaled from the DFT lattice parameter of 3.93 \AA~ to the experimental parameter of 3.87 \AA. Second, we include multiple rotations of the supercell tiled down the three $\left<100\right>$ directions to increase statistical sampling per atom column. The later is justified by the small displacements in the B sub-lattice. 

From the STEM simulations we determine the predicted NLN distances for the structure viewed along $\left<100\right>$ as a function of neighboring column intensity (Fig.~\ref{fig:figure2}c). To represent the atom column composition extremes, the measurements in Figure \ref{fig:figure2}c include STEM simulations of both the tiled structure and individual $\left<100\right>$ projections (Supplemental Material \cite{Sup_Mater} Figure 2). The predicted trends are consistent with the experimental observations (Fig.~\ref{fig:figure2}a and b), and the average NLN distance outliers are associated with pure Al or Ta atom columns from the original untiled DFT supercell. Inspection of the simulated tiled structure (Fig. \ref{fig:figure2}d), shows the same qualitative behavior as the experimental dataset (Figs.~\ref{fig:figure1} d and e) and predicts an A-A NLN standard deviation of  $\sim$ 10 pm comparable to the experiment, $\sim$ 7 pm. Also in agreement, the predicted B-B average NLN has a small standard deviation of $\sim$ 2 pm, similar to the $\sim$3 pm from experiment. The agreement indicates that when the average projected distortions from STEM are combined with DFT, it becomes possible to quantitatively connect local chemical fluctuations with distortion.

\begin{figure}[h!]
     \includegraphics[width = 3.5in]{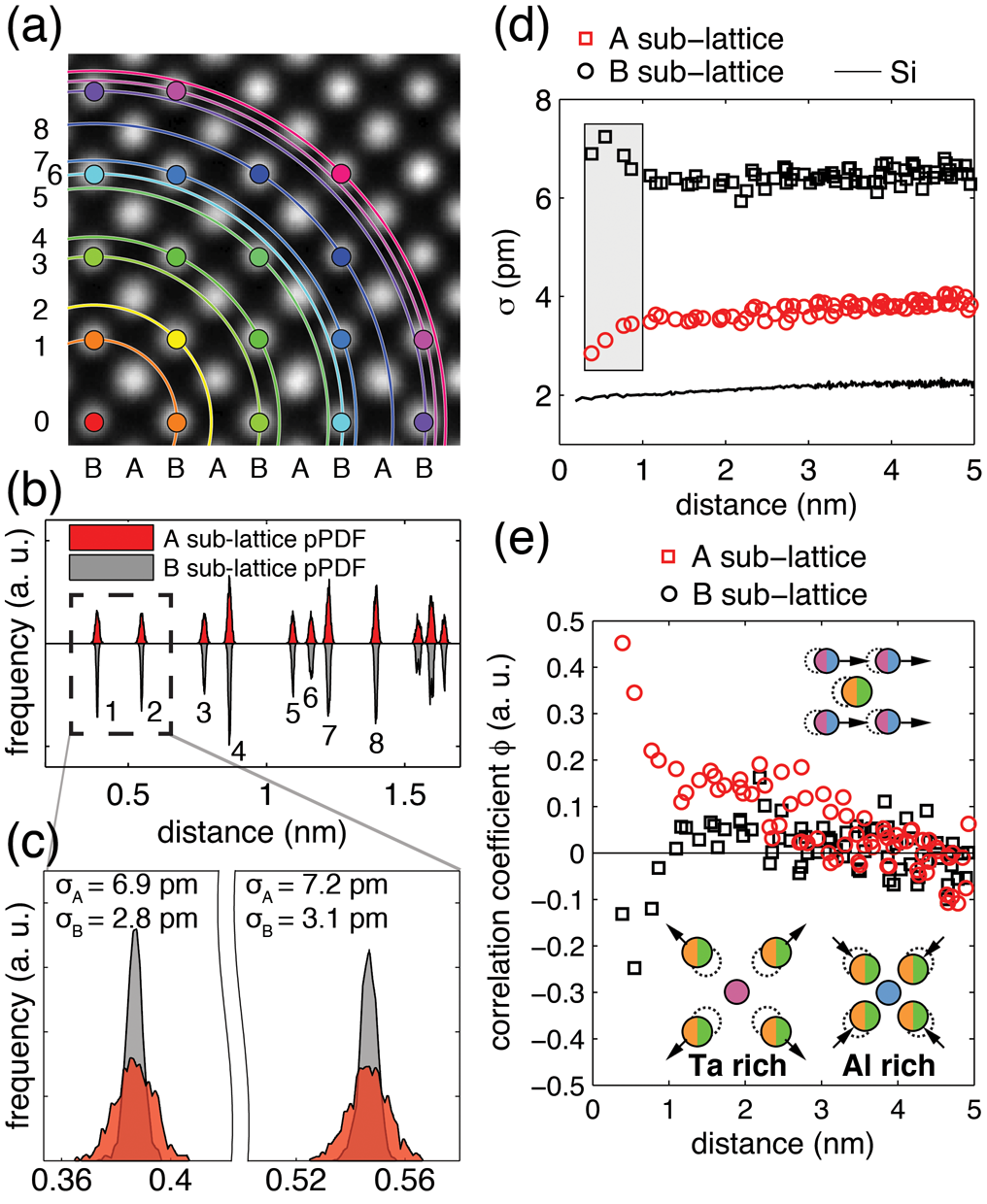}
  \caption{(a) Conceptual schematic of projected pair distribution function (pPDF).  (b) pPDFs for A  (red) and B (gray) sub-lattices calculated based on $n^{th}$ like-neighbor atom columns. (c) Comparison of first and second nearest like-neighbor peaks. (d) Standard deviation of the atom column pair distances for A (boxes) and B (cirlces) sub-lattices and Si (line). (e) Correlation coefficient $\phi$ as a function of $n^{th}$-neighbor distance for both sub-lattices. }
  \label{fig:figure3}
\end{figure}

To study structural correlations at longer length scales, $n^{th}$ like-neighbor distances are extracted across the entire image and used to construct a projected pair distribution function (pPDF).  This is schematically represented in Figure~\ref{fig:figure3}a for $1 \leq n \leq 11$. Conceptually, these measurements are similar to the pair distribution function (PDF) extracted from from diffraction data \cite{Petkov:2008aa}. The measurements differ, however, in that pPDFs provide insight into the projected average distortion, are determined directly at the atomic scale, and can be separated according to each sub-lattice.  While the pPDFs for A and B sub-lattices appear qualitatively similar (Fig.~\ref{fig:figure3}b), the standard deviation ($\sigma$) for the A-A nearest and second like-neighbors, as measured using the pPDF, are  considerably larger than those for the B-B distances and are reported in Figure \ref{fig:figure3}c. 

Extending the $\sigma$ analysis across the image, the A $n^{th}$ like-neighbor also differ dramatically from the B sub-lattice in terms of trend (Fig.~\ref{fig:figure3}d). Identical analysis was performed using a Si single crystal to demonstrate that the measurement precision is consistently $\approx$ 2 pm (Fig.~ \ref{fig:figure3}d, solid line). Inspection of the first few like-neighbors of LSAT reveals that $\sigma_B$ increases as the distance between B sub-lattice columns increases, while $\sigma_A$ increases initially and then decreases.  The stark difference between the LSAT trends, especially relative to the flat Si $\sigma$, indicates the presence of correlated atomic distortion within LSAT. To quantify the degree of correlation, we use the coefficient, $\phi = (\sigma_0^2-\sigma^2)/2\sigma_0^2$, where $\sigma_0$ is the contribution due to uncorrelated displacements at large pair distances.  This approach is commonly employed for PDF analysis \cite{Jeong:2003aa}, but without near perfect imaging enabled by RevSTEM, drift and noise would usually preclude extending precise distance measurements beyond a few angstroms (one unit-cell) in STEM.

As shown in Figure \ref{fig:figure3}e, the first few like-neighbors for the B sub-lattice exhibit strong, positive correlation while those for A show strong, negative correlation. The decreasing correlation for $\sigma_B$ can be explained by the rigid cage introduced by B-O covalent bonding: cooperative distortion results in near neighbor distances that are more similar than those at longer range (top schematic, Fig.~\ref{fig:figure3}e). Intriguingly, atypical negative correlation is observed for the A sub-lattice distortion.  This behavior can be understood in the context of the local chemical fluctuation of the B sub-lattice.  For Al- or Ta-rich atom columns, the surrounding A sub-lattice atoms contract inward or outward in opposite directions due to the local charge distribution (bottom schematic, Fig.~\ref{fig:figure3}e). Thus, A sub-lattice distortion is anti-correlated.  Furthermore, the second like-neighbor $\sigma_A$  increases due to compounding distortion contributions from both $x$ and $y$ directions (see further details of the correlation coefficient in the Supplemental Material \cite{Sup_Mater}). 

Through the power of picometer precise measurements in HAADF RevSTEM, we have demonstrated that relationships between chemistry and local distortion can be directly interrogated with atomic scale spatial resolution.  When combined with DFT, these results offer additional  insights to provide a complete mechanistic picture of unit-cell level structural displacements, which can be particularly important to the functionality of complex oxide solid solutions. The real-space measurement precision rivals that of diffraction determined PDFs, with the distinct advantage of  exposing a rich set of real space structural and chemical information.  This approach opens the doorway to new physical observations, as demonstrated here, and enable new opportunities to investigate property defining atomic structure.

\section*{Acknowledgments}

XS, EDG, and JML acknowledge the use of the Analytical Instrumentation Facility (AIF) at North Carolina State University, which is supported by the State of North Carolina and the National Science Foundation.  DLI acknowledges support from NSF Grant DMR-1151568 for this work. 

\section{}
\subsection{}
\subsubsection{}

\end{document}